\documentclass[conference]{IEEEtran}
\IEEEoverridecommandlockouts
\usepackage{cite}
\usepackage{amsmath,amssymb,amsfonts}
\usepackage{algorithmic}
\usepackage{graphicx}
\usepackage{textcomp}
\usepackage{xcolor}
\usepackage{url}
\usepackage{listings}
\usepackage{booktabs} 
\usepackage{multirow} 
\usepackage{float}
\usepackage{hyperref}
\usepackage{booktabs}
\usepackage{siunitx}
\usepackage{tabularx}
\usepackage{pgfplots}
\usepackage{fvextra} % extends verbatim
\pgfplotsset{compat=1.18}
\usepackage{tcolorbox}
\tcbuselibrary{skins}
\usepackage{fvextra} % for breaklines in Verbatim
\fvset{
  breaklines,
  breakanywhere,
  breaksymbolleft=,
  breaksymbolright=,
  breaksymbolindent=0pt
}
\tcbset{
  promptstyle/.style={
    enhanced,
    boxrule=0.5pt,
    colframe=black!40,
    colback=gray!4,
    arc=2pt,
    left=1mm,
    right=1mm,
    top=1mm,
    bottom=1mm,
    width=\columnwidth,
    before skip=5pt,
    after skip=5pt,
  }
}

\def\BibTeX{{\rm B\kern-.05em{\sc i\kern-.025em b}\kern-.08em
    T\kern-.1667em\lower.7ex\hbox{E}\kern-.125emX}}
\begin{document}

\title{Towards Agentic Investigation of Security Alerts
\thanks{© 2025 IEEE. Personal use of this material is permitted. Permission from IEEE must be obtained for all other uses, in any current or future media, including reprinting/republishing this material for advertising or promotional purposes, creating new collective works, for resale or redistribution to servers or lists, or reuse of any copyrighted component of this work in other works.}
\thanks{This research was funded by the European Union as part of the Horizon Europe project SYNAPSE (GA No. 101120853). Views and opinions expressed are, however, those of the author(s) only and do not necessarily reflect those of the European Union. Neither the European Union nor the granting authority can be held responsible for them.}\\

}
\author{\IEEEauthorblockN{Even Eilertsen}
\IEEEauthorblockA{\textit{University of Oslo} \\
Oslo, Norway \\
\texttt{eveneil@ifi.uio.no}}
\and
\IEEEauthorblockN{Vasileios Mavroeidis}
\IEEEauthorblockA{\textit{University of Oslo} \\
Oslo, Norway \\
\texttt{vasileim@ifi.uio.no}}
\and
\IEEEauthorblockN{Gudmund Grov}
\IEEEauthorblockA{\textit{Norwegian Defence Research Establishment (FFI)} \\ \& \textit{University of Oslo} \\
Kjeller, Norway \\
\texttt{Gudmund.Grov@ffi.no}}
}

\maketitle

\begin{abstract}
Security analysts are overwhelmed by the volume of alerts and the low context provided by many detection systems. Early-stage investigations typically require manual correlation across multiple log sources, a task that is usually time-consuming. In this paper, we present an experimental, agentic workflow that leverages large language models (LLMs) augmented with predefined queries and constrained tool access (structured SQL over Suricata logs and grep-based text search) to automate the first stages of alert investigation.
The proposed workflow integrates queries to provide an overview of the available data, and LLM components that selects which queries to use based on the overview results, extracts raw evidence from the query results, and delivers a final verdict of the alert. Our results demonstrate that the LLM-powered workflow can investigate log sources, plan an investigation, and produce a final verdict that has a significantly higher accuracy than a verdict produced by the same LLM without the proposed workflow. 

By recognizing the inherent limitations of directly applying LLMs to high-volume and unstructured data, we propose combining existing investigation practices of real-world analysts with a structured approach to leverage LLMs as virtual security analysts, thereby assisting and reducing the manual workload.
\end{abstract}


\section{Introduction}\label{sec:intro}
Modern Security Operation Centres (SOCs) are challenged by a large volume of alerts generated by their security tools \cite{tariq2025alert, nobles2022stress}. When security analysts face an overwhelming number of alerts, they are often able to investigate only a limited number of them. This phenomenon, commonly referred to as alert fatigue, leads to employee churn and increases the likelihood of genuine incidents being overlooked, consequently, diminishing the overall effectiveness of the SOC \cite{kearney2023combating}. As a result, most modern SOCs are exploring and adopting various forms of automation to assist analysts and lessen the manual workload associated with individual alerts, thereby enabling them to focus on more critical events \cite{brewer2019could}.

Artificial intelligence has played a significant role in cybersecurity for decades \cite{khraisat2019survey}. Identifying patterns and detecting threats based on anomalies is a field where machine learning has had great success. Recent advancements in Large Language Models (LLMs) have demonstrated strong performance in more generalised tasks and the ability to interact with other systems \cite{NEURIPS2023_d842425e}. LLMs can act as virtual security analysts and solve tasks similar to those that today's security analysts manually solve \cite{karlsen2024benchmarking}. The industry is rapidly developing products aimed at automating portions of the work performed by security analysts and advancing toward more autonomous AI agents that extend beyond the traditional Q\&A tasks of chatbots \cite {microsoft2025securitycopilot}.

In this paper, we examine the potential benefits of employing an LLM-based agentic approach to enhance and automate specific steps in security investigations. This paper focuses on testing agentic behaviour, where the LLMs have some autonomy in selecting actions and interpreting the results, but a more defined pipeline than a fully autonomous agentic system. Specifically, we assess how effectively LLMs can independently gather and synthesise information, and make informed decisions based on common security logs. The following research questions guide our study:
\begin{description}
    \item[\textbf{(RQ1)}] To what degree can the proposed agentic approach for alert investigation correctly query and perform operations on log sources to extract required data and contextualise alerts?
    \item[\textbf{(RQ2)}]To what degree can the proposed agentic approach accurately identify true and false positive alerts?
    \item[\textbf{(RQ3)}] How robust is the proposed approach to the stochastic variability inherent in large language model inference when processing identical input data?
\end{description} 
To address these research questions, we develop and evaluate a structured LLM-assisted workflow for security investigations, designed to address current limitations in the early stages of incident handling. The workflow consists of several coordinated LLMs that, albeit not fully autonomous, exhibit agentic behaviour enabling them to perform investigation tasks effectively.

The paper is structured as follows: in section \ref{sec:background} we provide some background material and discuss related work; in section \ref{sec:approach} we describe our approach; section \ref{sec:Methodology} describes our experiments with the results provided in section \ref{sec:results}; finally, we conclude and outline future work in section \ref{sec:conclusion}.

\section{Background}\label{sec:background}

LLMs are increasingly utilised in cybersecurity, with their task-solving capabilities showing promising performance in context-aware reasoning tasks within the security domain \cite{mohammed2025agentic}. A recent study investigating LLM applications in the security domain highlighted significant potential for automating tasks and personalised training, with challenges remaining in model generalisation, ethical deployment, and production readiness \cite{atlam2025llms}.
The ability to take text tokens and generalise knowledge is a function that can be utilised across many sub-fields of security \cite{ferrag2024generative}.

Context and prompt engineering is a growing field, especially in the applied usage of LLMs in systems that require a higher level of reliable performance\cite{white2024Patterns}. A comprehensive survey \cite{mei2025survey} of more than 1,400 research papers has shown that while current models augmented by context engineering demonstrate high proficiency in understanding complex context, some limitations lie in their ability to generate similarly complex outputs. The survey also investigated the evolution of context engineering, covering systems like \emph{Retrieval-Augmented Generation} (RAG) and more autonomous agentic systems like \emph{AutoGen}\footnote{\emph{AutoGen is an open-source programming framework created by  \hyperlink{https://www.microsoft.com/en-us/research/project/autogen/}{Microsoft} for building AI agents and facilitating cooperation among multiple agents to solve tasks.}} \cite{wu2024autogen}.
Multi-agent systems were not a primary focus; however, context engineering is stated to be an important part for success as these systems emerge. 
Whether it's full autonomy, tool calling, or deciding what lines of code to run, even smaller language models have performed well in zero-shot tool usage \cite{NEURIPS2023_d842425e}, where zero-shot entails that no training examples with the tools are provided in the prompt.
\newline\newline
Generative AI has seen massive and rapid adoption from both the offensive side and 
defensive security product providers \cite{microsoft2024digitaldefense}. According to the Splunk CISO 2025 report \cite{splunkCISOreport}, 65\% of CISOs are actively training security teams on prompt engineering, and 56\% are working on protocols for determining what tasks are more appropriate for AI or humans. They also focus on which tasks AI can assist humans with, especially in the cybersecurity industry. This development is advancing rapidly, with several agentic cybersecurity products emerging and continually updated as of 2025 \cite{microsoft2025securitycopilot, googleCloudSecurityAI, paloAltoAIcopilots}. Industry leaders, cybersecurity professionals, and academic researchers are increasingly recognizing the potential of applied generative AI in cybersecurity, driving innovation in the form of new approaches and products \cite{wef2024outlook, mohammed2025agentic, kshetri2025transforming}.

\section{LLM-powered Workflow for Alert Investigation}\label{sec:approach}

This section presents the approach, the architecture, and the implementation of our proposed workflow for LLM-assisted security alert investigation. At a high level, the workflow employs distinct LLM roles to investigate and synthesize evidence from available data before issuing a final verdict on the security alert. The design follows a modular architecture, enabling models, log types, queries, and tools to be easily replaced or customised to suit the needs of different adopters. By utilising techniques to improve LLM applications, such as context and prompt engineering, the workflow supports analysts in the initial stages of alert handling, particularly when available data is insufficient for decision-making. The section further details the individual components, their role, and the technologies employed in the implementation. All prompts and queries used in the experiments are available on GitHub\footnote {https://github.com/Rub3cula/CyberHunt2025/}.

\begin{figure}[H]
        \centering\includegraphics[width=0.8\linewidth]{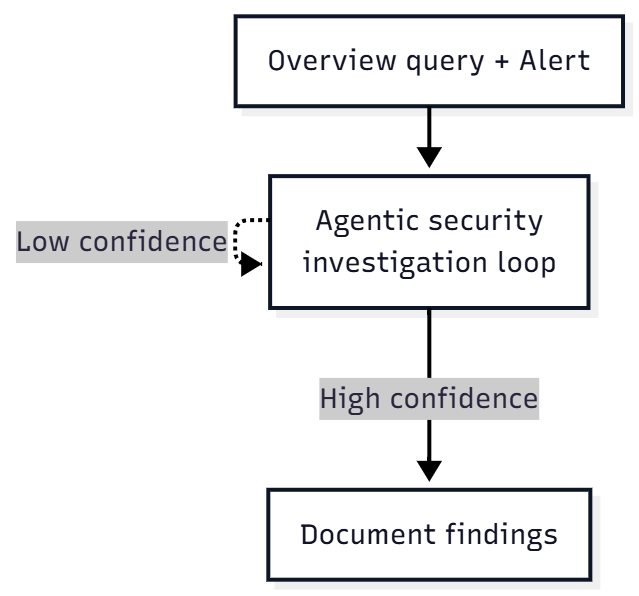}
        \caption{Investigation workflow} 
        \label{FullWorkFlow}
    \end{figure}

Figure \ref{FullWorkFlow} illustrates the overall architecture of the proposed LLM-based agentic workflow. The core of the system is the iterative \emph{agentic security investigation loop}, which is depicted in Figure \ref{FullWorkFlow}. To gather initial information as a starting point for the automated investigation, the first step is to query the available Suricata logs.\footnote {Suricata is a signature-based network Intrusion Detection System (IDS).}  
This query produces a summary containing total event counts, number of alerts and non-alerts, and top offending signatures (SIDs), source IPs, and destination IPs. This information is assumed to be necessary for most investigations, and we henceforth call this query
the \textbf{overview query}. The overview query is sent to the agentic security investigation loop, along with the original alert text.

\begin{figure}[H]
            \centering 
            \includegraphics[width=0.5\linewidth]{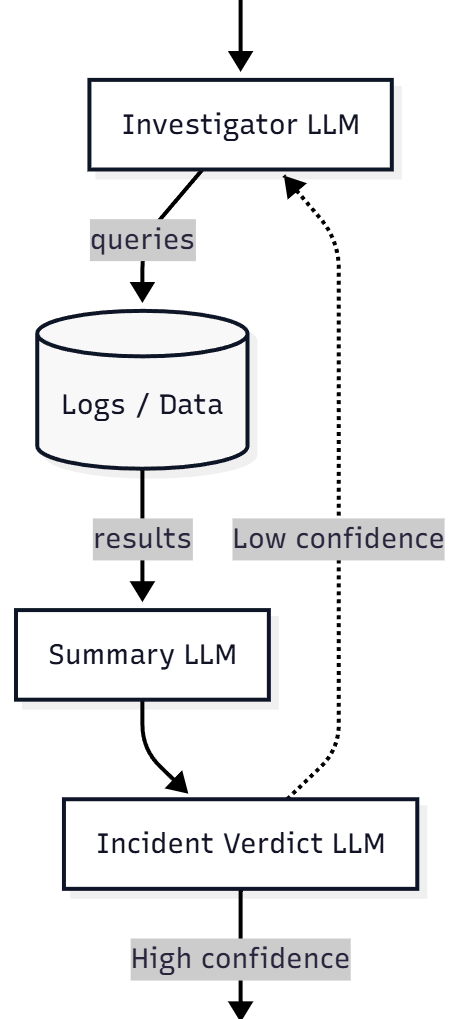}
            \caption{Agentic security investigation loop} 
            \label{AgenticLoop}
        \end{figure}

Figure \ref{AgenticLoop} expands on the \textit{Agentic security investigation loop} box from Figure \ref{FullWorkFlow} by illustrating the iterative steps that are performed by the different LLM components. The loop is designed to mimic the iterative process a security analyst would take, moving between querying data sources, analyzing results, and refining their search. Each step of the loop performs a part of the investigation to analyse the alert and associated log data as detailed below. The modular design allows for flexibility and the potential integration of additional LLM components or data sources in the future.

\paragraph{Investigator LLM}
The investigation loop starts with sending the alert, along with the overview query, to the \emph{Investigator LLM}. The Investigator is the first of the three LLM components\footnote{By different LLM components,' we mean different, distinct new calls to the LLM with different prompts.}, tasked with initiating the investigation and identifying relevant data for further analysis. In the prompt the LLM is provided with relevant information to provide context for its task and is tasked with performing a role similar to that of a security analyst, performing the initial step in an investigation. 

The Investigator LLM operates on both structured and unstructured data sources. Structured data comes in the form of Suricata IDS logs, which are readily queryable. These logs could be stored in a SQL database or, more realistically, in a SIEM solution such as Splunk or Elastic. For the remainder of the text, we assume a simple SQL database, as this is the approach used in our experiment, but this could easily be replaced with querying a SIEM solution.

The workflow also incorporates limited unstructured log data, which requires a different approach. To address this, we include a Linux \texttt{grep} operation as a fundamental querying mechanism for unstructured text. While not a formal database query, \texttt{grep} enables the LLM to search for specific patterns or keywords within unstructured logs, effectively serving as a basic query language for text-based data.

\paragraph{Summary LLM}
The third step is \emph{Summary LLM}. It uses all available data gathered: the initial alert details, the overview query, and the results from the last steps. It is then prompted to summarise all this data in an executive summary. If there are clear indicators that the logs contain evidence of a true positive or a false positive alert\footnote{As is common, the positive class is suspicious activity, and a \emph{true positive} is thus an alert that requires further analysis. In contrast, a \emph{false positive} is a false alert that can be ignored.}, it will present that data for the next step. It is important to note that while \emph{Summary LLM} is a critical component and often serves as the primary basis for the final verdict, it does not choose if its output is enough for a final judgment; however, its outputs will be the only source of information for the prompt to the next LLM, where a decision is made.

\paragraph{Incident Verdict LLM}
The fourth step is the \emph{Incident Verdict LLM}. The LLM takes the output from the \emph{Summary LLM}, and is tasked with and making a decision based on this data; is the available data enough to make a decision on the incident? The potential outcome from this step is:
\begin{enumerate}
    \item \emph{Malicious}: The Verdict LLM was presented with sufficient evidence to support that the alert is malicious (true positive). With this verdict, the loop completes, and the results are written to a file.
    \item \emph{Benign}: Enough evidence is presented to suggest that the alert is a false positive. This also completes the loop with the final verdict and results written to a results file.
    \item \emph{Uncertain}: Based on the provided evidence, there is no clear conclusion based on the set parameters, but no clues that could provide clear indicators if further investigated. This category could have been further divided or tuned to have a subset based on confidence, but it is judged as inconclusive for now, especially for benign cases.
    \item \emph{Requires Further Investigation (iterate)}: The evidence provided is inconclusive; additional data is needed to reach a definitive decision. The provided data will be returned to the \emph{Investigator LLM} to execute the loop another time. Currently, we only allow the loop to execute twice. This is to reduce complexity in the experiment. This constraint could be relaxed to allow multiple iterations, with each iteration performing a deeper investigation. This, however, requires a proper termination condition and a loop variant to ensure that the loop eventually terminates. 
\end{enumerate}

\section{Methodology}\label{sec:Methodology}

\subsection{Dataset}
To evaluate our approach, we use logs from two time windows from a the AIT Log Data Set V1.1 \cite{landauer2020aitlds}: The full dataset contains several attacks from several different endpoints, targeting different endpoints. According to the dataset's ground truth, there is a time window during the multi-stage attack in which a user performs several malicious actions including reconnaissance, brute-force and initial access attack on a web server. The first set of logs is extracted during the attack, which from here is referred to as the malicious or true positive subset. The benign logs referred to as the benign or false positive subset, is extracted from the same web server outside the active attack window, where it is assumed that all activities are benign, based on the ground truth from the full dataset.
\newline\newline
For the experiments, we defined a hypothetical triggered alert, assumed to be originating from a machine learning-based Intrusion Detection System (IDS). It is important to note that this alert is not found in the dataset, but it is assumed as the trigger that starts the investigation. Meaning that if the investigation concludes that there is no malicious traffic, this alert is a false positive, and true positive if malicious traffic is found. IDSes are well-suited for detecting anomalous patterns, such as sudden spikes in authentication attempts; however, they typically use considerably less information than a signature-based alert (e.g., from Suricata).\footnote{To illustrate, a signature-based alert could have information about a particular vulnerability (CVE) or an attack described in a threat report. On the other hand, a machine-learning-based IDS may only highlight an anomaly, such as increased traffic to an endpoint.} In this case, the IDS identified unusual traffic activity but provided no further context beyond the generic alert message "\emph{Suspicious behaviour}''.
\newline\newline
Following standard incident response practices, we extracted all log entries from the affected endpoint and the web server within a 30-minute window surrounding the triggered alert (25 minutes before and 5 minutes after). By limiting the experiment's scope to a specific phase of a multi-stage attack, the focus is on alert-level investigation rather than full incident analysis. Such a focus on alert-level investigation is particularly relevant for SOCs that have not implemented a state-of-the-art \emph{extended detection and response} (XDR) system, which is usually able to group alerts into incidents \cite{tariq2025alert}. Our more simplified approach enables evaluation of the workflow’s ability to investigate alerts generated by reliable detection mechanisms that deliver strong detection performance but limited contextual information. 

The subsets contain logs from a 30-minute period on the same web server, corresponding to a period without any attacks according to the ground-truth labels. The purpose was to investigate how the workflow handled false alerts, as the majority of IDS alerts tend to be false \cite{alahmadi202299}. Subsets are used to evaluate the workflow.

To emulate the diverse data landscapes commonly found in SOCs, we structured the subsets to include a mix data formats. The Suricata logs serve as detailed network event logs, processed into a SQL-compatible format using DuckDB, enabling efficient querying and data manipulation. The remaining logs, such as \emph{auth} logs and \emph{syslog}, were stored in a simple text format and serve as more unstructured data, which is not fully compatible with the SOC system that relies on Suricata's schema. Some operations can still be performed, such as using the \texttt{grep} command, to check for specific strings or frequencies within certain time windows. 
\newline\newline
These set of logs, even with its limitations, is sufficient to illustrate the type of investigation functionality conducted in a SOC. Although the implementation is rather crude, by using basic SQL and the grep command with some variables. Regardless the setup still provides the LLMs with necessary functionality. The logs also reflects the lack of real-world data in cybersecurity research, which is a well-known challenge, and this is particularly true when it comes to gathering realistic data for this type of SOC operations. Therefore, these steps in data selection and experiment setup aim to provide a somewhat realistic setting and address an important challenge for investigating security alerts in a production environment.

\subsection{Prerequisites} 

Before detailing the workflow, the prerequisites and underlying technologies used in the experiments need to be discussed.
Several experiments have been conducted, and each experiment utilizes only one version of the four selected LLMs. If GPT-5-mini is used in a run, the LLM component throughout that entire workflow will consistently prompt the same model. 
\begin{itemize}
\item \textit{GPT-5-mini} by OpenAI (accessed through API);
 \item \textit{claude-3-haiku-20240307} from Anthropic (accessed through API);
 \item \textit{qwen3:30b} from Alibaba (executed locally); 
 \item \emph{gemma3:27b} from Google (executed locally).
\end{itemize}
 These models were selected for their diversity and popularity of adoption. During the early stages of research, these cheaper, smaller models demonstrated adequate performance, enabling more frequent testing runs. A complete execution of 100 runs for each model costs less than 10 USD for models accessed via API (\textit{GPT-5-mini} and \textit{claude-3-haiku-20240307}), while the local models can execute 100 runs in less than 1 hour on a 4090 NVIDIA GPU. \newline\newline This setup enabled a comparative analysis of the performance of popular LLMs and cost-effective reproduction or implementation. 

As analytical reasoning is a crucial part of the workflow, some data is needed to quantitatively evaluate the inputs and outputs of \textit{Summary LLM} and \emph{Incident Verdict LLM}. These data points are stored as artifacts alongside the complete response and prompt of the prior LLM components. Inputs to the last step of the workflow are a culmination of the results of the prior steps, meaning that the prompt to \emph{Incident Verdict LLM} provides insights into the results from the two other LLMs. By analysing this output during the evaluation process, we can gain insights into which inputs led to specific conclusions. For future work, this analysis is crucial for tuning the system using techniques such as prompt engineering. 

Before the investigation, the Suricata logs from the web server had been processed to enable SQL queries to be executed.

\subsection{Workflow}
In this subsection we illustrate our workflow as shown in Figure \ref{AgenticLoop}, with a concrete example.
\paragraph{Step 1} The first step starts with the invocation of \emph{Investigator LLM}. If it is the first run, \emph{Investigator LLM} will receive the alert text and the results from the overview query; if not, the output from \emph{Incident Verdict LLM} is also part of the input to \emph{Investigator LLM}. The LLM is prompted to propose a plan to collect evidence related to the behaviours identified in the input. It is provided with a selection of four queries from which it can choose a subset (including applying all of them). Additionally, it may include a custom freeform SQL query where the query is not fully defined, but the LLM completes it. The investigator is also instructed to include a \texttt{grep} operation that searches the logs for keywords such as ``\texttt{failure}" and other relevant keywords. The \texttt{grep} operation is dynamic, meaning the LLM is free to find suitable keywords, but it is also provided with examples it can use.

A simplified system and user prompt for the \emph{Investigator LLM} are displayed below:

\begin{tcolorbox}[promptstyle,title={\textbf{Investigator LLM System Prompt}}]
\begin{Verbatim}[breaklines,breakanywhere,fontsize=\scriptsize]
You are part of the Automated Security SOC workflow.
We are investigating a potential incident on a web server.
You will receive access to structured Suricata IDS logs
and limited unstructured logs 
You are a single-step LLM component: you work only with the data given.
Your responsibility is to propose a compact plan to collect high-signal evidence that can later be mapped to MITRE ATT&CK behaviors.\n"
    

ROLE:
You are the Investigator.
Do not fabricate data.

BEHAVIORAL CONTEXT (write behaviors, not technique numbers):\n"
    - Credential abuse: repeated login failures, source IP concentration, username spraying.
    - Exploitation of public-facing apps: exploit-like strings, malformed HTTP methods, sensitive admin/auth paths, repeated attempts.
    - Webshell/persistence indicators: uploads that create executable server-side files.
    - Scanning and reconnaissance: broad probing, fuzzed methods/paths, rapid clustered activity

PLAN REQUIREMENTS:
- Choose up to 4 queries from the predefined queries
- Optionally ONE 'free_sql'
- ALWAYS include a GREP object that YOU define:
\end{Verbatim}
\end{tcolorbox}

\begin{tcolorbox}[promptstyle,title={\textbf{Investigator LLM Prompt}}]
\begin{Verbatim}[breaklines,breakanywhere,fontsize=\scriptsize]
You receive results from a query to give insights summarizing some Suricata alerts.
Your job: output a COMPACT plan to collect more evidence.
Choose up to 4 queries from the allowed set.
Optionally add ONE 'free_sql'
ALWAYS include 'grep' with YOUR chosen regex alternation in 'keywords'

Rules:
Each query must specify a limit of results
Keep 'summary' equal or less than 3 lines
\end{Verbatim}
\end{tcolorbox}

Using the alert text and the overview query, the \emph{Investigator LLM} selects up to four predefined queries, one optional custom SQL (freeform), and a mandatory \texttt{grep} operation to gather additional evidence relevant to the alert. The choices of SQL and how to populate the regex and grep are not clearly defined; only some guidance is provided on where to start. Below are the available choices for the Investigator LLM:

\emph{Predefined queries and operations:}
\begin{itemize}
  \item \texttt{sids\_window} \emph{(Top alert signatures in window)}. 
  Purpose: rank the most frequent Suricata signatures to establish dominant alert themes. Returns: \texttt{sid}, representative \texttt{msg} \texttt{(ANY\_VALUE)}, \texttt{max\_sev}, \texttt{ct}.
  \item \texttt{top\_src\_alerts} \emph{(Top source IPs generating alerts)}. Purpose: identify prolific alerting sources. Returns: \texttt{src\_ip}, \texttt{ct}.
  \item \texttt{top\_dst\_alerts} \emph{(Top destination IPs targeted by alerts)}. Purpose: surface primary victim hosts. Returns: \texttt{dest\_ip}, \texttt{ct}.
  \item \texttt{http\_paths\_alerts} \emph{(Alerted HTTP paths with sample status)}. Purpose: summarize HTTP paths involved in alerting flows. Returns: \texttt{http\_path}, \texttt{ct}, example \texttt{http\_status} \texttt{(ANY\_VALUE)}.
  \item \texttt{freeform\_regex} \emph{(Regex filter over alert messages)}. Purpose: targeted pattern-driven slice across \texttt{msg} for exploit strings or auth markers. Returns: \texttt{ts}, \texttt{src\_ip}, \texttt{dest\_ip}, \texttt{sid}, \texttt{severity}, \texttt{msg} ordered by time.
  \item \texttt{grep} \emph{(Semi-structured logs)}. Purpose: keyword or regex search over auth or syslog text files in the time window. Returns: file path, matched line samples, occurrence counts.
\end{itemize}

\paragraph{Step 2} The execution step takes the plan from \emph{Investigator LLM} and performs the actions in the plan. The selected queries are executed with some guardrails: all SQL is time-bounded (\texttt{ts BETWEEN ? AND ?}), parameterized (placeholders only, no string interpolation), and hard-limited by a caller-supplied limit in the range [1, 5]. Regex patterns for \texttt{freeform\_regex} and \texttt{grep} are validated; the freeform SQL must be a single \texttt{SELECT} over the \texttt{suricata} table, include a \texttt{LIMIT} that is normalized, and exclude mutating or unsafe keywords. Some per-run metrics, such as \texttt{syntax\_ok}, \texttt{rows}, \texttt{ms}, \texttt{nonempty}, and \texttt{grep\_success}, are stored as artefacts for troubleshooting and to increase some insight into the workflow. The results of the query are also stored as artefacts for the following steps and the final verdict. With these guardrails, the predefined queries are sanitised and executed, with output limits to prevent unnecessary data from being stored or passed along in the workflow. The results of the queries and \texttt{grep} command are stored as JSON objects.

The following example is a simplified output from \emph{Investigator LLM}, showcasing the names of four predefined queries, one query with a custom regular expression, and a generated query and some of its rationale.
\begin{tcolorbox}[promptstyle,title={\textbf{Shortened Investigator LLM example output}}]
\begin{Verbatim}[breaklines,breakanywhere,fontsize=\scriptsize]
"queries": [
    {
      "name": "sids_window",
      "name": "top_src_alerts",
      "name": "top_dst_alerts",
      "name": "timeline_alerts",
      
      "name": "freeform_regex",
        "params": {
            "pattern": "pass=|password|upload|shell|cmd=|/admin|/cfide|/servlet|php",
        
  "free_sql": {
    "sql": "SELECT ts, src_ip, dest_ip, proto, sid, severity, msg, http_method, http_path, http_status, host FROM suricata WHERE ts BETWEEN ? AND ? ORDER BY ts LIMIT ?",
  },
  "rationale_bullets": [
    "freeform_regex: hunt for plaintext credentials, upload attempts, web shells or cmd patterns in HTTP payloads.",
    "free_sql: quick sample of raw records to validate fields and collect artifacts (paths, status codes) for follow-up."
  ]
}</final>
\end{Verbatim}
\end{tcolorbox}

\paragraph{Step 3} \emph{Summary LLM} receives the results from the execution step. It is tasked with building a behaviour-focused executive summary from the input and computing a confidence score between 0 and 1, where 1 represents full confidence. There are several strict instructions in the prompt regarding the format and the requirement not to invent data (hallucination). The confidence score is just an output from the model that serves to guide the model in decision-making based on its own output. To guide and tune the confidence score, there are several instructions and examples of what to look for. A simplified, extracted part of the prompt is displayed below

\begin{tcolorbox}[promptstyle,title={\textbf{Summary LLM Prompt}}]
\begin{Verbatim}[breaklines,breakanywhere,fontsize=\scriptsize]
You are the SUMMARY LLM
    Your job is to build a executive summary and compute confidence regarding if the data provided should be considered malicious or benign based on the observed evidence.
    STRICT EVIDENCE RULES:
    - Never invent data.
    - Query samples can hint at suspicious paths/messages but must not be fabricated.
\end{Verbatim}
\end{tcolorbox}

The output from this step includes an executive summary in plain language, as well as any Indicators of Compromise (IoC) or MITRE ATT\&CK techniques.
The output is stored as a JSON object. An example of a shortened output:
\begin{tcolorbox}[promptstyle,title={\textbf{Shortened Summary LLM example output}}]
\begin{Verbatim}[breaklines,breakanywhere,fontsize=\scriptsize]
"executive_summary": "The investigation reveals indicators of potential credential abuse and unauthorized access attempts on the web server. The evidence shows a high volume of failed login attempts from a single IP address, as well as repeated attempts to access a sensitive login page. While no definitive webshell or exploitation activity was observed, the abnormal authentication patterns and probing of sensitive areas warrant further investigation to determine the scope and nature of the potential incident."
"iocs": {"ips": ["127.0.0.1"], "paths": ["/login.php"]}, "mitre": [{"technique": "Credential Access", "confidence": 0.9, "why": "The evidence shows a high number of failed login attempts (286) from a single IP address (127.0.0.1), indicating potential credential abuse or probing."},{"technique": "Active scanning / malformed HTTP probing", "confidence": 0.5, "why": "Suricata HTTP method anomaly (sid 2221030) with high alert counts and a top source generating thousands of events suggests broad/malformed HTTP scanning or fuzzing activity"}
auth_signal: fail_lines: 286, top_users: daryl, 286]
\end{Verbatim}
\end{tcolorbox}

\paragraph{Step 4} \emph{Incident Verdict LLM} is tasked with assigning a verdict to the alert, which could be benign, malicious or uncertain. It receives the same instructions as the other LLMs, explaining the role, input, and output. It also receives some guidance with examples of cases where the verdict would be malicious or benign. For the experiments, the examples are pretty basic and has minor details, but still provides a adequate overview of potential attacks. \emph{Incident Verdict LLM} is instructed to be strict with the malicious or the benign, as these should only be used if the confidence is high, but uncertain can be assigned if unsure. When the evidence is inconclusive but suggests potential malicious activity, the LLM may assign a verdict of “uncertain” and recommend further investigation. This allows security analysts to prioritize alerts requiring deeper review, rather than dismissing potentially critical signals. The following example presents a shortened output from \emph{Incident Verdict LLM}, including the final verdict and relevant information, along with suggested next steps.

\begin{tcolorbox}[promptstyle,title={\textbf{Shortened Verdict LLM example output}}]
\begin{Verbatim}[breaklines,breakanywhere,fontsize=\scriptsize]
 "verdict":"malicious",
  "confidence":0.9,
  "mitre":[
    {
      "technique":"Credential abuse / brute force",
      "confidence":0.9,
      "why":"286 failed authentication attempts concentrated to a single user ('daryl') and single source (127.0.0.1) indicates high-volume, focused brute-force activity against local services."
    },
    {
      "technique":"Active scanning / malformed HTTP probing",
      "confidence":0.7,
      "why":"Suricata HTTP method anomaly (sid 2221030) with high alert counts and a top source generating thousands of events suggests broad/malformed HTTP scanning or fuzzing activity."
    }
  "next_steps":
    "Credential response: immediately reset/expire the 'daryl' account password and any shared/related credentials; enforce or enable MFA for affected accounts.",
\end{Verbatim}
\end{tcolorbox}

\paragraph{Step 5} When an investigation has received the final verdict by \emph{Incident Verdict LLM}, the results and some relevant metrics from the run are written to a file for performance evaluation and troubleshooting. Key metrics captured include the run\_id for tracking individual runs, the verdict assigned (benign, malicious, or uncertain), and the associated confidence score. The identified mitre\_techniques are also included if any is found, along with a count of these techniques (mitre\_count), providing insights into the types of behaviors detected. Technical details like the number of queries planned (planned\_count), successful executions (syntax\_ok\_count), and results returned (nonempty\_count) are also recorded, allowing for optimization of the query process. These metrics, as detailed in Table \ref{tab:incident-metrics} provides more insight into each investigation and is important to our evaluation of the LLM’s performance, troubleshooting and tuning.

\begin{table}
\centering
\footnotesize
\setlength{\tabcolsep}{6pt}
\renewcommand{\arraystretch}{1.1}
\begin{tabularx}{\columnwidth}{@{} l l X @{}}
\toprule
\textbf{Field} & \textbf{Type} & \textbf{Description} \\
\midrule
run\_id               & str         & Unique run key. \\
run\_label            & str         & Human label; `.1` suffix = reiterate. \\
model                 & str         & Model ID (grouping variable). \\
verdict               & enum        & \{benign, malicious, uncertain\}. \\
confidence            & float [0,1] & Decision strength (0-1). \\
mitre\_techniques     & str         & Semicolon-separated list of behavior labels. \\
mitre\_details        & str         & \texttt{label=score} pairs (for quick inspection). \\
mitre\_count          & int         & Number of distinct behavior labels. \\
planned\_count        & int         & Number of queries planned. \\
syntax\_ok\_count     & int         & Queries that executed without error. \\
nonempty\_count       & int         & Queries that returned one or more rows. \\
free\_sql\_syntax\_ok & \{0,1\}     & Free-SQL validated and ran. \\
free\_sql\_nonempty   & \{0,1\}     & Free-SQL returned one or more rows. \\
grep\_ran             & \{0,1\}     & GREP executed (files found). \\
grep\_success         & \{0,1\}     & GREP found relevant lines of code. \\
\bottomrule
\end{tabularx}
\vspace{3pt}
\caption{Per-run metrics.}
\label{tab:incident-metrics}
\end{table}

\subsection{Baseline approach} 

\begin{figure}[H]
            \centering 
            \includegraphics[width=0.45\linewidth]{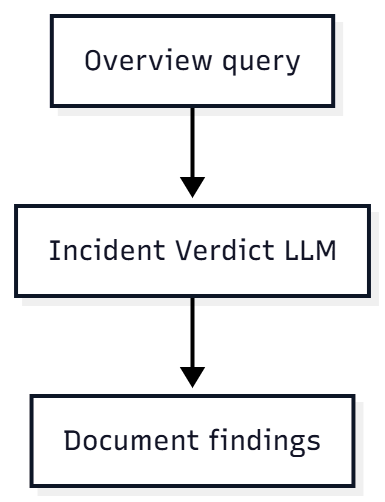}
            \caption{Workflow for baseline experiment} 
            \label{BaseLine}
        \end{figure}

To evaluate our proposed agentic approach -- including its need and ability to enrich the alert with additional context in the form of logs -- we have implemented a simple baseline approach without such enrichment. This baseline approach is shown in Figure \ref{BaseLine}. It is established by directly submitting the overview query and the alert description to \emph{Incident Verdict LLM} without the preceding steps. Minimal changes are made from what \emph{Investigator LLM} receives, and the prompt for the \emph{Incident Verdict LLM} is identical. This direct approach serves as the baseline performance of \emph{Incident Verdict LLM} without the workflow's automated investigation and enrichment. The baseline experiment was repeated 100 times on all models on both subsets.

\section{Results}\label{sec:results}

In this section, we present the outcomes of the experiments. The workflow has been tested on both subsets, with each model executed 100 times. Throughout this evaluation, accuracy is used as the percentage of runs where the final verdict (Malicious, Benign, or Uncertain) correctly matches the ground truth label. For each subset and model, accuracy is calculated by dividing the number of correctly classified runs by the total number of runs. A run is considered correctly classified if the predicted verdict matches the actual label (i.e., a malicious run is correctly identified as malicious, and a benign run is correctly identified as benign). The overall accuracy is then the average of the accuracies calculated for each individual subset.

\subsection{Baseline approach}

The verdict distribution of the baseline approach is shown in Table \ref{tab:verdict-dist-nopipeline}. The baseline configuration presents the Verdict LLM with the same initial information as the workflow (overview query + alert text). Notably, the baseline achieved 0\% accuracy on the malicious subset, with all models incorrectly classifying malicious runs as benign. The Gemma model consistently predicted benign verdicts across both subsets. Further investigation revealed that, even with varied input data, Gemma invariably returned a benign verdict, indicating a strong bias towards classifying all inputs as non-malicious. While Gemma correctly identified benign runs, the justification provided in its output was often empty or lacked any nuanced analysis.

\begin{table}
\centering
\scriptsize
\setlength{\tabcolsep}{4pt}
\renewcommand{\arraystretch}{1.0}
\begin{tabular}{ll S[table-format=3.0] S[table-format=3.0] S[table-format=3.0]}
\toprule
\textbf{Subset} & \textbf{Model} & {\textbf{Mal. (\%)}} & {\textbf{Ben. (\%)}} & {\textbf{Unc. (\%)}} \\
\midrule
\multicolumn{5}{l}{\emph{Malicious}}\\
& Claude~3~Haiku & 0 & 1 & 99 \\
& Gemma~3:27B    & 0 & 100 & 0 \\
& GPT-5-mini     & 0 & 0 & 100 \\
& Qwen3:30B      & 0 & 35 & 65 \\
\midrule
\multicolumn{5}{l}{\emph{Benign}}\\
& Claude~3~Haiku & 0 & 14 & 86 \\
& Gemma~3:27B    & 0 & 100 & 0 \\
& GPT-5-mini     & 0 & 0 & 100 \\
& Qwen3:30B      & 0 & 63 & 37 \\
\bottomrule
\end{tabular}
\vspace{3pt}
\caption{Verdict distribution without the workflow (Baseline). Mal = Malicious, Ben = Benign, Unc = Uncertain}
\label{tab:verdict-dist-nopipeline}
\end{table}

\subsection{Agentic Workflow}

As shown in Table \ref{tab:verdict-dist-effective}, the proposed workflow resulted in strong performance for the malicious subset across all models, with significantly higher accuracy than the baseline. Three out of four models achieved above 90\% accuracy in the final verdict, with the exception of Gemma, which achieved 81\% accuracy. GPT-5-mini had zero mistakes in the final verdict, correctly classifying all runs as malicious. While some variations exist, all models performed very well, with a maximum false negative rate of only 4\%. 

\begin{table}[H]
\centering
\scriptsize
\setlength{\tabcolsep}{4pt}
\renewcommand{\arraystretch}{1.05}
\begin{tabular}{ll S[table-format=3.0] S[table-format=3.0] S[table-format=3.0]}
\toprule
\textbf{Subset} & \textbf{Model} & {\textbf{Mal. (\%)}} & {\textbf{Ben. (\%)}} & {\textbf{Unc. (\%)}} \\
\midrule
\multicolumn{5}{l}{\emph{Malicious}}\\
& Claude~3~Haiku & 94 & 0 & 6 \\
& Gemma~3:27B    & 81 & 4 & 15 \\
& GPT-5-mini     & 100 & 0 & 0 \\
& Qwen3:30B      & 95 & 3 & 2 \\
\midrule
\multicolumn{5}{l}{\emph{Benign}}\\
& Claude~3~Haiku & 1 & 89 & 10 \\
& Gemma~3:27B    & 0 & 52 & 48 \\
& GPT-5-mini     & 0 & 0 & 100 \\
& Qwen3:30B      & 1 & 85 & 14 \\
\bottomrule
\end{tabular}
\vspace{3pt}
\caption{Verdict distribution by model. Mal = Malicious, Ben = Benign, Unc = Uncertain}
\label{tab:verdict-dist-effective}
\end{table}

\emph{GPT-5-mini} demonstrated the strongest performance on the malicious subset, with 100 percent correct verdict on the malicious data. However, it it also consistently classifying the benign data as uncertain, even while utilising the same prompt as the other models. This behaviour does indicate over-conservatism given the same prompt as the others. Nevertheless all models tested on the benign dataset had a exceptionally low rate of incorrectly malicious verdict, with only 2 out of 400. 

Table \ref{tab:reiterate-usage} illustrates the frequency of repeated evaluations during the 100 workflow executions on each model for each subset. These iterations indicate instances where \emph{Incident Verdict LLM} requested additional information before reaching a verdict. The considerable variation in iteration frequency across models suggests differences in their behavioural approaches, consistent with the observed results in uncertain verdicts in both the baseline and workflow experiment.
\begin{table}[H]
\centering
\scriptsize
\setlength{\tabcolsep}{5pt}
\renewcommand{\arraystretch}{0.95}
\begin{tabular}{ll S[table-format=3.0] S[table-format=3.0] S}
\toprule
\textbf{Subset} & \textbf{Model} & {\textbf{Effective Runs}} & {\textbf{Iterations}} & {\textbf{Iterations (\%)}} \\
\midrule
\multicolumn{5}{l}{\emph{Malicious}}\\
& Claude~3~Haiku & 100 & 15 & 15.0 \\
& Gemma~3:27B    & 100 & 37 & 37.0 \\
& GPT-5-mini     & 100 & 0 & 0.0 \\
& Qwen3:30B      & 100 & 0 & 0.0 \\
\midrule
\multicolumn{5}{l}{\emph{Benign}}\\
& Claude~3~Haiku & 100 & 75 & 75.0 \\
& Gemma~3:27B    & 100 & 35 & 35.0 \\
& GPT-5-mini     & 100 & 98 & 98.0 \\
& Qwen3:30B      & 100 & 0 & 0.0 \\
\bottomrule
\end{tabular}
\vspace{3pt}
\caption{Amount of reiterations.}
\label{tab:reiterate-usage}
\end{table}

\subsection{Analysing the Results}

Overall, the results demonstrate that the proposed workflow successfully 
achieved high accuracy on the final verdict on true positive alert, while being more conservative and careful in verdicts on the false positive alert. 

The models differ substantially even when using identical prompts. While some of the parts of the workflow are static, there is considerable variety and scope for tuning. Specifically, the prompts have significant potential for tuning to improve performance. 

As shown in Table IV, the results demonstrate the value of incorporating the workflow to enhance the security investigation. Regardless of limitations due to the simplicity of the designs, the subset, or the scope of the investigation, the workflow delivers a significant improvement in verdict accuracy compared to the baseline experiment without it.

Beyond improvements in verdict accuracy, the LLM responses show significant potential to enrich security investigations beyond the metrics presented in the tables. An example of useful output can be seen in a response from a \emph{Summary LLM} run, where it wrote in the executive summary that the verdict should be malicious, and it included a field stating \begin{verbatim}
"why": "286 authentication failures for 
user 'daryl' from 127.0.0.1 strongly 
suggests credential abuse attempts."
\end{verbatim} While there are many of these results in the work, it is mentioned in the future work as promising work within understanding the results and giving some insights into the justifications of the LLM.

\section{Conclusion}\label{sec:conclusion}

We have proposed an agentic security investigation workflow and compared it with a deliberately simple baseline workshop to demonstrate the feasibility and usefulness of an agentic approach. We use basic prompts and context engineering to address simple attacks, such as brute-force, and assume an anomaly-based alert with limited contextual information to highlight the need to build sufficient context from logs, a crucial step in alert analysis. Our results show promise, with the following analysis of our proposed research questions:

\paragraph{RQ1} The LLMs demonstrated strong capabilities in leveraging the predefined queries, retrieving relevant information from the log sources, and dynamically creating new queries with the given instructions. The queries and limitations in the experiments were relatively narrow. Still, compared with the baseline experiments, our approach showed that it could both extract necessary log data and understand the logs to make a correct verdict.

\paragraph{RQ2} The workflow demonstrates high accuracy in identifying malicious behaviour in the log data. In particular, the malicious detection has high accuracy, with an average model performance of $93\%$. \emph{GPT-5-mini} indicated a strong performance with 100\% accuracy. 

For the benign subset, some models demonstrated a preference to assign uncertainty, Fewer than 1\% of the workflow runs on the benign subset had malicious verdicts. We argue that prioritizing the elimination of incorrect verdicts is more valuable than minimizing uncertainty, as a well-performing model can handle a portion of the alerts, freeing up analyst time. While some uncertainty remains, and necessitates manual investigation, this is preferable to incorrectly classifying malicious activity as benign. 
 The workflow demonstrated a significant improvement over the simpler baseline.

\paragraph{RQ3} Certain aspects of the workflow have been hard-coded. The code for the execution uses the same logic for each run. The variability comes from each LLM. The inherent characteristic of LLMs results in output variation. That variation affects certain aspects downstream in the workflow, but with the added sanitation and instructions in the prompts, the results remain relatively consistent across runs, as shown in the results. 

To successfully build automation for security investigations, achieving a robust workflow that has high accuracy in detecting true positives while minimizing false negatives is crucial. With minimal false negatives, the potential risk of implementing such a system is lower, and LLMs can be used to save time. The saved time can then be spent on catching the edge cases and manual verification of results, and further improving security.
If the objective of implementing LLMs is to assist or save time rather than fully replace analysts, then enriching alerts and creating some automation on the initial steps in an investigation is a suitable use case. Therefore, minimizing incorrect verdicts, even at the potential cost of an increased number of uncertainty verdicts, directly supports our goal of automating alert triage and reducing the burden on security analysts.

\paragraph{Concluding remarks}

While we believe the results demonstrate promising potential, there are significant opportunities to refine LLM behaviour, optimise prompts, and improve overall implementation. This workflow is meant as a proof-of-concept to lay some groundwork for future exploration rather than a production-ready solution.
The empirical findings of this work are that the proposed simple implementation of an LLM-powered investigation workflow automates some steps of investigation or assists in investigating a security alert. The models behave differently even when prompted with the same prompt. We conclude by suggesting some promising next steps.

\subsection{Future Work and limitations}

Our initial evaluation utilised a limited data representing a single attack scenario with a narrow, imagined alert. We believe this is sufficient for a preliminary assessment, but acknowledge it fails to capture the full complexity and diversity of real-world security logs and data, which should be addressed in the future.
Below, we discuss some avenues for future work which we believe have considerable potential and could address some of the limitations. 

\paragraph{Real world security data} Future exploration should incorporate more extensive and diverse datasets. This data should include a variety of attacks, logs and system configurations. Moreover, the alerts should be from real IDS systems, including both machine learning-based and signature-based, and there should be greater diversity in the enrichment of contextual data, such as MITRE ATT\&CK techniques and Common Vulnerabilities and Exposures (CVEs).

\paragraph{Extensive tuning} Due to the non-determinism in LLMs, there should be a significant amount of experimentation with different prompts and tuning of parameters, both inside and outside of the prompts.

\paragraph{Dynamic thresholds} Dynamic thresholds could be implemented, and based on other factors, the threshold for the final verdict can be moved up and down. For example, if an organisation knows it is benignly targeted, it could reduce the threshold to be more aggressive in issuing malicious verdicts, and during a quieter period, revert to a lower threshold. 

\paragraph{Specialised Agents} The LLMs in this implementation exhibited some agentic behaviour, but cannot be seen as proper AI agents. More fully autonomous agents could be a part of a similar workflow. One example of a quick feature to add to the workflow would be the ability for the workflow to change the timewindow These agents could be specialised into domains such as reverse engineering, SQL query builder, or email protection. Combined, this could build an agentic SOC that is partly or fully autonomous for certain security scenarios.

\paragraph{Security considerations and adversarial robustness} There are inherent security risks with implementing LLMs into such systems. Like many existing systems, LLMs are susceptible to various attacks, including prompt injection and other manipulation techniques. This needs to be addressed.


\begin{thebibliography}{00}

\bibitem{tariq2025alert}
S. Tariq, M. Baruwal Chhetri, S. Nepal, and C. Paris, 
“Alert fatigue in security operations centres: Research challenges and opportunities,” 
\textit{ACM Computing Surveys}, vol. 57, no. 9, pp. 1–38, Apr. 2025. 
[Online]. Available: https://dl.acm.org/doi/pdf/10.1145/3723158[Accessed: Sep. 7, 2025].

\bibitem{nobles2022stress}
C. Nobles, 
“Stress, burnout, and security fatigue in cybersecurity: A human factors problem,” 
\emph{Holistica Journal of Business and Public Administration}, 
vol. 13, no. 1, pp. 49--72, 2022. 
[Online]. Available: \url{https://sciendo.com/pdf/10.2478/hjbpa-2022-0003} 
[Accessed: Oct. 10, 2025].

\bibitem{kearney2023combating}
P. Kearney, M. Abdelsamea, X. Schmoor, F. Shah, and I. Vickers, 
“Combating alert fatigue in the security operations centre,” 
\textit{SSRN Electronic Journal}, Nov. 15, 2023. [Online]. Available: \url{https://papers.ssrn.com/sol3/papers.cfm?abstract_id=4633965} [Accessed: Sep. 7, 2025].

\bibitem{brewer2019could} R. Brewer, “Could SOAR save skills-short SOCs?,” \textit{Computer Fraud \& Security}, vol. 2019, no. 10, pp. 8–11, 2019. [Online]. Available: https://doi.org/10.1016/S1361-3723(19)30106-X [Accessed: Sep. 7, 2025].

\bibitem{khraisat2019survey}
A. Khraisat, I. Gondal, P. Vamplew, and J. Kamruzzaman,
“Survey of intrusion detection systems: techniques, datasets and challenges,”
in \emph{Cybersecurity}, 
vol. 2, no. 1, pp. 1--22, 
Springer, 2019. 
[Online]. Available: \url{https://cybersecurity.springeropen.com/articles/10.1186/s42400-019-0038-7}


\bibitem{NEURIPS2023_d842425e}
T. Schick, J. Dwivedi-Yu, R. Dessì, R. Raileanu, M. Lomeli, E. Hambro, L. Zettlemoyer, N. Cancedda, and T. Scialom, 
“Toolformer: Language models can teach themselves to use tools,” 
in \emph{Advances in Neural Information Processing Systems}, 
vol. 36, 
A. Oh, T. Naumann, A. Globerson, K. Saenko, M. Hardt, and S. Levine, Eds. 
Red Hook, NY, USA: Curran Associates, Inc., 2023, 
pp. 68539--68551. 
[Online]. Available: \url{https://proceedings.neurips.cc/paper_files/paper/2023/file/d842425e4bf79ba039352da0f658a906-Paper-Conference.pdf}

\bibitem{karlsen2024benchmarking}
E. Karlsen, X. Luo, N. Zincir-Heywood, and M. Heywood, “Benchmarking large language models for log analysis, security, and interpretation,” Journal of Network and Systems Management, vol. 32, no. 3, p. 59, 2024. [Online]. Available: https://link.springer.com/article/10.1007/s10922-024-09831-x
. [Accessed: Sep. 7, 2025].

\bibitem{microsoft2025securitycopilot}
Microsoft, “Microsoft Security Copilot,” Microsoft Security, 2025. [Online]. Available: https://www.microsoft.com/en-us/security/business/ai-machine-learning/microsoft-security-copilot. [Accessed: Sep. 7, 2025].

\bibitem{mohammed2025agentic}
A. Mohammed, 
“Agentic AI as a proactive cybercrime sentinel: Detecting and deterring social engineering attacks,” 
\emph{Journal of Data and Digital Innovation (JDDI)}, 
vol. 2, no. 2, pp. 109--117, 2025. 
[Online]. Available: \url{https://datalensjourna.com/index.php/JDDI/article/view/18/13} 
[Accessed: Oct. 10, 2025].

\bibitem{atlam2025llms}
H. F. Atlam, “LLMs in Cyber Security: Bridging practice and education,” Big Data and Cognitive Computing, vol. 9, no. 7, p. 184, 2025. [Online]. Available: https://doi.org/10.3390/bdcc9070184
. [Accessed: Sep. 7, 2025].

\bibitem{ferrag2024generative}
M. A. Ferrag, F. Alwahedi, A. Battah, B. Cherif, A. Mechri, and N. Tihanyi, “Generative AI and large language models for cyber security: All insights you need,” SSRN, 2024. [Online]. Available: https://ssrn.com/abstract=4853709
. [Accessed: Sep. 7, 2025].

\bibitem{white2024Patterns}
J. White, S. Hays, Q. Fu, J. Spencer-Smith, and D. C. Schmidt,
“ChatGPT prompt patterns for improving code quality, refactoring, requirements elicitation, and software design,”
in *Generative AI for Effective Software Development*, 
A. Nguyen-Duc, P. Abrahamsson, and F. Khomh, Eds.
Cham: Springer Nature Switzerland, 2024, pp. 71--108.
[Online]. Available: \url{https://doi.org/10.1007/978-3-031-55642-5_4}
[Accessed: Oct. 28, 2025].


\bibitem{mei2025survey}
L. Mei, J. Yao, Y. Ge, Y. Wang, B. Bi, Y. Cai, J. Liu, M. Li, Z.-Z. Li, D. Zhang, \emph{et al.}, 
“A survey of context engineering for large language models,” 
\emph{arXiv preprint arXiv:2507.13334}, 2025. 
[Online]. Available: \url{https://arxiv.org/abs/2507.13334} 
[Accessed: Oct. 10, 2025].

\bibitem{wu2024autogen}
Q. Wu, G. Bansal, J. Zhang, Y. Wu, B. Li, E. Zhu, L. Jiang, X. Zhang, S. Zhang, J. Liu, \emph{et al.}, 
“AutoGen: Enabling next-gen LLM applications via multi-agent conversations,” 
in \emph{Proceedings of the First Conference on Language Modeling}, 
2024. 
[Online]. Available: \url{https://arxiv.org/abs/2308.08155} 
[Accessed: Oct. 10, 2025].

\bibitem{microsoft2024digitaldefense}
Microsoft, “Microsoft Digital Defense Report 2024,” Microsoft Security Insider, 2024.  
[Online]. Available: \url{https://www.microsoft.com/en-us/security/security-insider/intelligence-reports/microsoft-digital-defense-report-2024}  
[Accessed: Oct. 10, 2025].

\bibitem{splunkCISOreport}
Splunk, “The CISO Report,” Splunk, 2025.  
[Online]. Available: \url{https://www.splunk.com/en_us/form/ciso-report/}  
[Accessed: Oct. 10, 2025].

\bibitem{googleCloudSecurityAI}
Google, “Security + AI on Google Cloud,” Google Cloud, 2024.  
[Online]. Available: \url{https://cloud.google.com/security/ai}  
[Accessed: Oct. 10, 2025].

\bibitem{paloAltoAIcopilots}
Palo Alto Networks, “AI Copilots Simplify Cybersecurity,” Palo Alto Networks, 2024.  
[Online]. Available: \url{https://www.paloaltonetworks.com/precision-ai-security/copilots}  
[Accessed: Oct. 10, 2025].

\bibitem{wef2024outlook}
World Economic Forum  Accenture, “Global Cybersecurity Outlook 2024,” World Economic Forum, Jan. 2024.  
[Online]. Available: \url{https://www3.weforum.org/docs/WEF_Global_Cybersecurity_Outlook_2024.pdf}  
[Accessed: Oct. 10, 2025].

\bibitem{kshetri2025transforming}
N. Kshetri, 
“Transforming cybersecurity with agentic AI to combat emerging cyber threats,” 
\emph{Telecommunications Policy}, 
p. 102976, 2025. 
[Online]. Available: \url{https://doi.org/10.1016/j.telpol.2025.102976} 
[Accessed: Oct. 10, 2025].

\bibitem{landauer2020aitlds}
M.~Landauer, F.~Skopik, M.~Wurzenberger, W.~Hotwagner, and A.~Rauber,  
\emph{AIT Log Data Set V1.1},  
Zenodo, 2020.  
DOI:10.5281/zenodo.4264796.  
[Online]. Available: \url{https://zenodo.org/records/4264796}  
[Accessed: Oct. 11, 2025].  

\bibitem{alahmadi202299}
B.~A.~Alahmadi, L.~Axon, and I.~Martinovic,
``99\% false positives: A qualitative study of SOC analysts' perspectives on security alarms,''
in \textit{Proceedings of the 31st USENIX Security Symposium (USENIX Security 22)},
2022, pp.~2783--2800.

\end{thebibliography}
\end{document}